\documentclass{article}
\usepackage{spconf,amsmath,graphicx}
\usepackage{graphicx}
\usepackage{amsmath}

\usepackage{booktabs}
\usepackage{framed}
\usepackage{color}
\usepackage{soul}
\usepackage{subfigure}
\usepackage{verbatim}
\usepackage{pifont}
\usepackage{stmaryrd}

\usepackage[ruled,noline,noend,linesnumbered]{algorithm2e}

\title{Privacy-preserving Cloud-based DNN Inference}
\name{Shangyu Xie, Bingyu Liu and Yuan Hong}
\address{Illinois Institute of Technology\\\{sxie14, bliu40\}@hawk.iit.edu, yuan.hong@iit.edu}
\begin{document}
%
\maketitle
\begin{abstract}
Deep learning as a service (DLaaS) has been intensively studied to facilitate the wider deployment of the emerging deep learning applications. However, DLaaS may compromise the privacy of both clients and cloud servers. Although some privacy preserving deep neural network (DNN) techniques have been proposed by composing cryptographic primitives, the challenges on computational efficiency have not been fully addressed due to the complexity of DNN models and expensive cryptographic primitives. In this paper, we propose a novel privacy preserving cloud-based DNN inference framework (``PROUD''), which greatly improves the computational efficiency. Finally, we conduct experiments on two datasets to validate the effectiveness and efficiency for the PROUD while benchmarking with the state-of-the-art techniques. 

\end{abstract}

\section{Introduction}\label{sec:intro}

Deep neural network (DNN) models have been frequently deployed in a wide variety of real world applications, such as image classification \cite{liu2017survey}, video recognition \cite{tran2015learning} and voice assistant (e.g., Apple Siri and Google Assistant). Meanwhile, cloud computing technologies (e.g., Microsoft Azure Machine Learning, Google Inference API, and Amazon AWS Machine Learning) have promoted the deep learning as a service (DLaaS) to make DNNs widely accessible. Users can outsource their own data for inferences based on the pre-trained DNN models provided by the cloud service provider.

However, severe privacy concerns may arise in such applications. First, if the data of the clients are explicitly disclosed to the cloud, sensitive personal information included in the outsourced data would be leaked. Second, if the fine-tuned DNN models are shared for inferences \cite{jiang2018secure}, the parameters might be reconstructed by untrusted parties \cite{tramer2016stealing}. To address such privacy concerns, several recent works \cite{juvekar2018gazelle,gilad2016cryptonets,xie2019bayhenn,zhang2018gelu} have proposed cryptographic protocols to ensure privacy in inferences via garbled circuits \cite{Yao86} and/or homomorphic encryption \cite{paillier1999public}), which rely on expensive cryptographic primitives. Then, such protocols may result in fairly high computation and communication overheads. Since the volume of the outsourced data grows rapidly and the DNN models usually require high computational resources in the cloud, such techniques may not be suitable for practical deployment due to limited scalability. Thus, we are seeking an efficient scheme to securely implement the DNN inferences in the cloud.

Specifically, we propose a privacy-preserving cloud-based DNN inference framework (``PROUD'') by co-designing the cryptographic primitives, deep learning, and cloud computing technologies. We mainly take advantage of a novel matrix permutation with ciphertext packing and parallelization to improve the computational efficiency of linear layers. With the privacy guarantee provided via homomorphic encryption, PROUD supports all types of non-linear activation functions by leveraging an interactive paradigm. Above all, PROUD integrates the cloud container technology to further improve the performance via parallel execution, which can also be readily adapted for various DNNs via configuring container images.

\section{The PROUD System}
\label{sec:system}

\begin{figure}[!h]
	\centering
	\includegraphics[angle=0, scale=0.65]{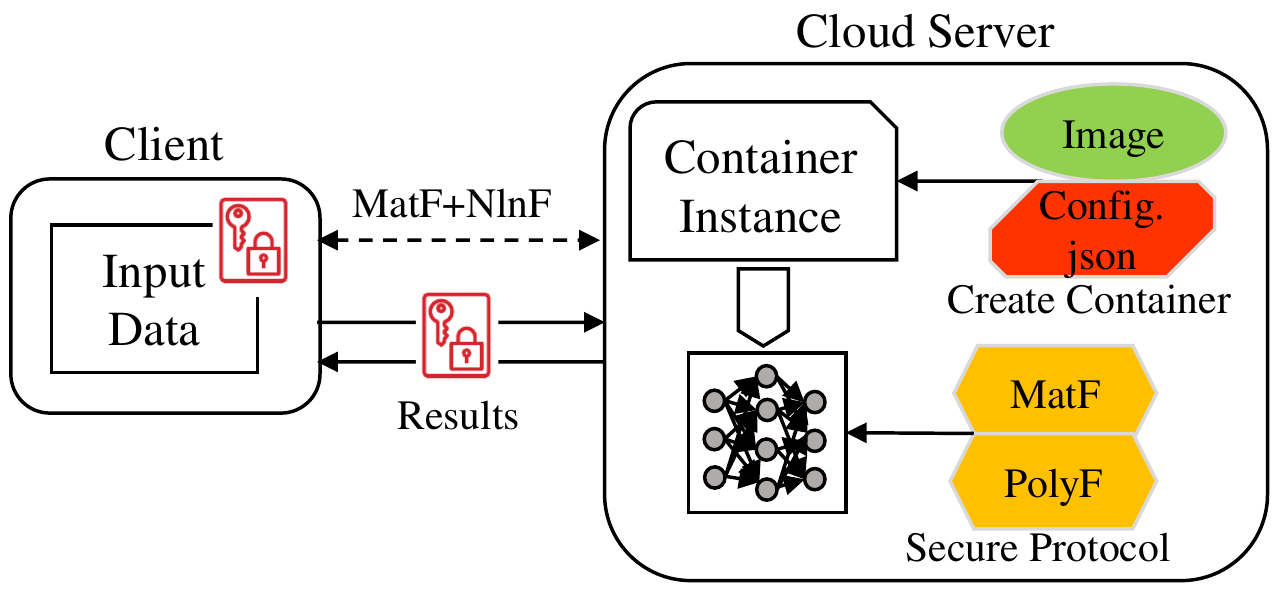}	\vspace{-0.05in}
	\caption[Optional caption for list of figures]
	{The PROUD Framework. }\vspace{-0.15in}
	\label{fig:framework}
\end{figure}

Figure \ref{fig:framework} illustrates the framework of the proposed system for the users (clients) and the cloud service provider (cloud server). The client locally holds the private data, which will be encrypted with the client's public key and sent to the cloud server. Then, the cloud server initializes container instances (pre-compiled with secure protocols, i.e., MatF and NlnF) to execute the DNN inference with the encrypted input. Finally, the client will decrypt and receive the classification result.

\vspace{0.05in}
\noindent\textbf{Automated Backend Execution}. The backend system can automatically deploy the cryptographic protocol for the secure data inference in the cloud. Specifically, once the server receives encrypted data from the client, it will compose the configuration file to initialize a bunch of container instances via a pre-compiled image (with the source codes), where the secure protocols (i.e., MatF and NlnF) will start to be executed for DNN inference until the final result is returned. The automation of the backend ensures that the secure protocols can be delivered efficiently, and enables the full system to be capable of processing a large number of clients (if necessary).

\section{The PROUD Protocol Design}
\label{sec:frame}

\subsection{Problem Formulation}

The PROUD will securely compute the DNN model with encrypted inputs in the cloud. We first denote an $\ell$-layer DNN model as $\mathcal{M}=\{L_i, i\in [1, \ell]\}$, and the input video as $\mathcal{V}$. The inference model $\mathcal{M}$ can be viewed as a complex function $f(\cdot)$ integrating linear functions (corresponding to linear layers, e.g., convolution layers and fully-connected layers) and non-linear functions (activation functions, e.g., Sigmoid and ReLu). Denoting the inference result as $\mathcal{S}$, we have: 
\begin{equation}
    \mathcal{S}=f(\mathcal{V})=L_\ell(L_{{\ell}-1}(\cdots L_2(L_1(\mathcal{V}))\cdots))
    \label{eq:ml}
\end{equation}

\noindent\textbf{Threat Model}. We consider \emph{semi-honest} model where both parties are honest to execute the protocol but are curious to learn private information. PROUD can preserve privacy for both parties against possible leakage: (1) client's private input videos are not leaked to the cloud service provider; (2) cloud service provider's DNN model (e.g., linear/non-linear weight parameters, and bias values) is not revealed to the client in the computation. We also assume that all the communications are executed in a secure and authenticated channel.

\subsection{Protocol Overview}

Algorithm \ref{algm:framework} illustrates the protocol for PROUD. In the initialization phase, the client generates a key pair and encrypts the private data $\mathcal{V}$ (Line 1) while the server prepares the computation for the DNN functions (Equation \ref{eq:ml}) with two subprotocols: (1) MatF for the linear functions; (2) NlnF for the non-linear activation functions (Line 2). With such two subprotocols, PROUD will be jointly executed by both the client and server. Specifically, the server can perform computation of the linear layers directly on the encrypted data received from the client using the subprotocol MatF (Line 5). For the non-linear layers, the output data will be sent back to the client for computation by the subprotocol NlnF (Line 6), and then the client will re-encode and encrypt the data to be sent to the server for next layer's computation. Once completing the computations of all the layers in the DNN model, the client will receive the ciphertext and decrypt it to get the classification result. The details of two subprotocols will be illustrated in Section \ref{sec:mat} and \ref{sec:poly}, respectively.

\begin{algorithm}
\footnotesize

\KwIn{Input Data $\mathcal{V}$, $\mathcal{M}$}

\KwOut{Classification Result $\mathcal{S}$}

Client: Encode and encrypt $\mathcal{V}$ to get $\tau_0$

Server: (MatF, NlnF) $\gets \mathcal{M}$

\For {$i \in [1, \ell]$ }{

\Switch{$L_i$}{
            \textbf{Case}~\emph{Linear}:~~$\tau_i\leftarrow \text{MatF}(\tau_{i-1})$
    
          \textbf{Case}~\emph{Non-Linear}:~~{$\tau_i\leftarrow \text{NlnF}(\tau_{i-1})$}
        }

}

Client: Decrypts $\mathcal{\tau_\ell}$ to get $\mathcal{S}$

\caption{PROUD Protocol}\label{algm:framework}

\end{algorithm}

\subsection{MatF Protocol} \label{sec:mat}
To ensure privacy for the linear layers, a naive method is to apply homomorphic encryption (HE) to the arithmetic operations of encrypted matrices (e.g., fully-connected layer), which might be inefficient since the input data tensors are usually high-dimensional. To mitigate such issue, our PROUD system utilizes a novel matrix permutation method \cite{jiang2018secure} to efficiently perform matrix computations with ciphertext packing and parallelization \cite{cheon2017homomorphic}, where the matrix multiplication equals the sum of the component-wise products for some specific permutations of the matrices themselves.

Given the input matrix $V$, the linear layer (matrix) $W$ and bias parameter $B$, PROUD will securely compute the function of a linear layer as: $W*V+B$ (w.l.o.g., we consider the fully-connected layer with bias while $W$ and $V$ are two square matrices with size $n \times n$). We illustrate an example of the square matrix as $A$ (of size $n\times n$). To compute the multiplication, the server will first find $n$ permutations of the matrix $A$ via the following symmetric permutations:

\begin{equation}
\small
    \sigma(A)_{i,j}=A_{i, i+j}, \tau(A)_{i,j}=A_{i+j,j}
\end{equation}
\begin{equation}
    \phi(A)_{i,j}=A_{i,j+1}, \psi(A)_{i,j}=A_{i+1,j}
\end{equation} 

Note that $\phi, \psi$ are the column and row shifting operations. Then, we can compute the product for $W$ and $V$ as below: \begin{equation}
\small
    W*V=\sum_{k=0}^{n-1} W_k \odot V_k
    \label{eq:component}
\end{equation}

where $W_k=\phi^k (\sigma(W))$, $V_k=\psi^k(\tau(B))$, $\odot$ indicates the component-wise product and $k$ is the number of perturbations, e.g., $\psi^k$ will perform $k$ times $\psi(\cdot)$ permutation on the matrix. We denote the function \emph{permut}($\cdot$) to compute the $n$ permutation matrices of one matrix.

\vspace{0.05in}

\noindent\textbf{Ciphertext Packing and Parallelization}. To improve the efficiency, we also leverage the vectorable homomorphic encryption (aka. ``Ciphertext Packing'') \cite{jiang2018secure,juvekar2018gazelle}, which transforms a matrix of size $d\times d$ to a single vector (plaintext) via an encoding map function, denoted as \textit{Encode}. In particular, the \textit{Decode} function transforms the vector plaintext back to the matrix form. For simplicity of notations, we denote the encryption, evaluation, and decryption functions under an HE scheme as $Enc(), Eval()$ and $Dec()$, respectively. 
 
Then, the component-wise product (Equation \ref{eq:component}) of the ciphertexts $V_k$ and $W_k$, denote as $Enc(pk,O_k)$, can be securely computed with the multiplicative property of the HE:

\begin{align}
\small
Eval(pk, Encode(W_k^{(l,m)}),Enc(Encode(V_k^{(l,m)})), *)
\end{align}
where $l, m \in[1,n]$ are the entry indices of the matrices $W$ and $V$, and $pk$ is the public key. Then, the sum of all the $n$ component-wise products of the matrices $W_k$ and $V_k$ can be computed using the additive property of HE. Finally, the bias parameter $B$ can be computed using the additive property of HE. The protocol is detailed in Algorithm \ref{algm:matrix}. 

Given a large number of plaintexts to be encrypted by ciphertext packing, we further expedite the matrix computation with the parallelization \cite{jiang2018secure}. To this end, we modify the encoding map function to ``1-to-1 map'' such that an $n$-dimensional vector can be transformed into a $g$-tuple of square matrices of order $d$, where $g=n/d^2$. This parallelization technique can also be realized with the parallel computation in the cloud framework (using a bunch of containers), which results in a reduced computational complexity $O(d/g)$ per matrix.

\begin{algorithm}
\footnotesize
\KwIn{Input $V$, Weighted Matrix $W$, Bias $B$}

\KwOut{$O=Enc(pk, W*V+B)$}

$\{V_k\}_{k=0}^{n-1} \gets Enc(pk,Encode(permut(V)))$

$\{W_k\}_{k=0}^{n-1} \gets Encode(permut(W))$

\For{$k \in [0,n-1]$}{

$O_k\leftarrow Eval(pk, W_k^{(l,m)},V_k^{(l,m)}), *)$

}

$Enc(pk, O)$ $\leftarrow$ $Eval(pk, \{O_k, k\in[0, n-1]\}, +)$

\Return{$Eval(pk,Enc(pk,O), B, +)$}
\caption{MatF}\label{algm:matrix}
\end{algorithm}

\subsection{NlnF Protocol} \label{sec:poly}

The NlnF protocol securely computes the non-linear layers of DNNs. Most of the existing works depend on either garbled circuits \cite{juvekar2018gazelle} or replacing square function \cite{jiang2018secure}, which may arouse high computational overheads or reduce the accuracy. In our protocol, the computation of the non-linear function (e.g., ReLu) is executed at the client side with the input of decrypted data to preserve privacy. Algorithm \ref{algm:poly} shows that the client will first decrypt the received output of MatF from the server with its private key. Then, the client will compute the output of the non-linear function $\phi$ and return the output to the server for the computation of next network layer. During the execution of this protocol, the client does not leak any private information to the server and the server does not expose sensitive weight parameters to the client.

\begin{algorithm}
\footnotesize
\KwIn{Input $V$ (from MatF), Activation Function $\phi(\cdot)$}

\KwOut{$O$}
=
Server: sends $V$ to the client

Client: $r\leftarrow Decode(Dec(sk, V))$

\Return{$O\leftarrow \phi(r)$}
\caption{NlnF}\label{algm:poly}
\end{algorithm}

\vspace{0.05in}

\noindent\textbf{Security and Practicality}. For the linear computations (MatF), the server will not know the plaintext since all the computations are performed on the ciphertexts (``no leakage'' can be theoretically proven). For the non-linear computations, the client receives some encrypted intermediate results from the server, and decrypts them to get some trivial intermediate data (which does not result in privacy leakage). Such trivial non-private data release is traded for a light-weight cryptographic protocol, which is far more efficient than other cryptographic protocols built on secure polynomial approximation and/or garbled circuits. Since the protocol is composed independently, many neural network based applications (e.g., image classification \cite{liu2017survey} and natural language processing \cite{devlin2018bert}) or video learning models (e.g., C3D \cite{tran2015learning} and I3D \cite{carreira2017quo}) can be readily integrated into our system. The pre-trained DNNs can be adapted with appropriate extensions, and integrated into the PROUD protocol (for feature extraction and/or inferences on the encrypted data). Moreover, the PROUD system can be easily integrated into the practical cloud platform (e.g., AWS) since the PROUD is a cloud-based prototype of system.

\section{Experiments}
\label{sec:exp}

\vspace{0.05in}
\noindent\textbf{Experimental Setup}. Our system is implemented on the NSF CloudLab platform\footnote{https://www.cloudlab.us/} in which one machine works as the client and the other one works as the server. Both machines have eight 64-bit ARMv8 cores with 2.4GHZ, 64GB memory installed with Ubuntu 16.04. We implement the homomorphic encryption in HEANN \cite{cheon2017homomorphic} (which realizes the optimal computation over real numbers) for secure matrix operations. We leverage Docker to develop the prototype for PROUD: the image of the container (all the source codes) is pre-compiled with the specific functions (i.e., MatF and NlnF) in Python. 
 
We evaluate our framework on the two datasets: (1) MNIST dataset \cite{lecun1998gradient} includes 70K handwritten images of size $28\times 28$ under the gray level 0-255; (2) IDC dataset\footnote{http://www.andrewjanowczyk.com/use-case-6-invasive-ductal-carcinoma-idc-segmentation/} for invasive ductal carcinoma
(IDC) classification (IDC-negative or positive), which contains about 28K patches of $50\times 50$ pixels. We employ the LeNet5 \cite{lecun1998gradient} as the test network model. In addition, we compare the performance of PROUD with four representative schemes (CryptoNets \cite{gilad2016cryptonets}, GAZELLE \cite{juvekar2018gazelle}, BAYHENN \cite{xie2019bayhenn} and DELPHI \cite{mishra2020delphi}) on the MNIST and IDC dataset for image classification.

\begin{table}
\footnotesize
\centering
\begin{tabular}{cccc}  
\toprule
Framework  & Accuracy (\%) & Latency (s) & Comm. (MB)\\
\midrule
CryptoNets       & 96.09  & 1080.3   & 595.5        \\
GAZELLE    & 99.05  & 8.05 & 100.65     \\
BAYHENN   & 98.93  & 2.34  & 20.81   \\
DELPHI   & 96.2  & 0.84  & 0.81   \\
PROUD  & 99.01  & 0.62 & 1.03          \\
\bottomrule
\end{tabular}
\caption{Benchmarking on MNIST dataset}\vspace{-0.1in}
\label{tab:mini}
\end{table}
\vspace{-0.05in}
\begin{table}
\footnotesize
\centering
\begin{tabular}{cccc}  
\toprule
Framework  & Accuracy (\%) & Latency (s) & Comm. (MB)\\
\midrule
CryptoNets       & 81.25  & 1942.6   & 1621.3        \\
GAZELLE    & 83.74  & 24.64 & 263.4   \\
BAYHENN   & 83.26  & 9.36  & 67.32   \\
DELPHI   & 82.72  & 2.47  & 2.95   \\
PROUD  & 84.01  & 1.12 & 3.27         \\
\bottomrule
\end{tabular}
\caption{Benchmarking on IDC dataset}\vspace{-0.2in}
\label{tab:idc}
\end{table}

\vspace{0.05in}
\noindent\textbf{Results}. All the results on the two datasets are shown in Table \ref{tab:mini} and \ref{tab:idc}, respectively. From the Table \ref{tab:mini}, we can observe that our PROUD results in the least average latency (e.g., 13 times faster than GAZELLE) and communication overheads for digit classification, compared with other three existing schemes. PROUD significantly outperforms other schemes considering we adopt a highly light-weight matrix computation scheme compared with the existing schemes (including garbled circuits and heavily encrypting matrices). As for the classification accuracy, PROUD works almost identical as GAZELLE (in which the optimal approximation of non-linear function achieves the negligible loss using the original activation function). It is worth noting that CryptoNets performs the worst, since it replaces all the activation functions with the square functions, and all the pooling functions with sum pooling, which also greatly increase the computational overhead and arouse the high communication bandwidth (the larger ciphertext size). BAYHENN uses a different Bayesian inference model with some randomness for DNN, which decreases the classification accuracy to some extent. Also, considering that the DELPHI's computation overheads are mainly in the offline preparation (heavy cryptographic computations), the online computation overhead is reduced. Table \ref{tab:idc} shows similar results for IDC classification. All of these results illustrate that our proposed framework can significantly improve the computational efficiency of secure DNN inference compared with other state-of-the-art techniques. 

We also illustrate the evaluation results of latency and communication bandwidth result for each step of PROUD processing one image instance in Table \ref{tab:mnist_sys}. Specifically, the client takes about 23.4 ms, including the runtime for encoding and encrypting the image. Then, the server initializes the DNN model by taking 107.2 ms (note that this step can be processed simultaneously as the first step at the client's). Moreover, we also observe that DNN computation in the server dominates the latency. Regarding the communication overheads, it mainly occurs when the client sends the encrypted images to the server (0.58MB). Moreover, there arouses communication consumption during the DNN inference since NlnF protocol follows an interactive paradigm.

\begin{table}
\footnotesize
\centering
\begin{tabular}{c|ccc}  
\toprule
  & Phase  & Latency (ms) & Comm. (MB)\\
\midrule
Client   & Encode + Encry.  & 23.4    & 0.58        \\
Server    & Set Model  & 107.2 & -      \\ 
Server   & DNN Computation  & 410.8   & 0.34    \\
Client  &Decry.+ Decode  & 2.7  & 0.03      \\
\hline
  &Total  & 544.1 &0.95         \\
\bottomrule
\end{tabular}
\caption{Performance of PROUD on MNIST dataset}
\label{tab:mnist_sys}
\end{table}

\section{Related Work}
\label{sec:related}
The generic secure computation techniques (e.g., secure two-party computation \cite{Yao86,goldreich1987play}, fully homomorphic encryption \cite{Gentry:FHE} and secret sharing \cite{boyle2017homomorphic}) can be directly used to tackle the privacy concerns in DNN inferences. However, such cryptographic primitives would request high computation and communication overheads. For instance, the size of garbled circuits in the MPC protocols will exponentially grow as the number of parties increases. They also require multiple rounds of communications among the parties. Recently, although there are multiple works that improve the efficiency of FHE \cite{cryptoeprint:2012:144,halevi2018faster,halevi2019improved}, the high computational costs still make them impractical for performing inferences.

Therefore, it seems to be necessary to design specific protocols for secure learning. There have been several works on designing specific secure protocols for DNN models \cite{gilad2016cryptonets,mohassel2017secureml,juvekar2018gazelle,xie2019bayhenn}. SecureML \cite{mohassel2017secureml} is one of the first systems which focuses on machine learning on encrypted data with NN model. However, it mainly depends on the generic two-party protocols with very poor performance. 
Jiang et al. \cite{jiang2018secure} proposed an efficient secure matrix computation protocol to improve the performance for the computation with neural networks. However, it only supports limited activation functions (e.g., only the square function in the case study). GAZELLE \cite{juvekar2018gazelle} composes the protocol on the heavy garbled circuits while BAYHENN \cite{xie2019bayhenn} leverages inefficient ciphertext packing of matrix for linear computations. Although DELPHI improves the bandwidth of online protocol, it still depends on the off-line computation and neural architecture search (NAS).

\section{Conclusion}
\label{sec:concl}

We have proposed a novel privacy preserving DNN inference framework with cloud container technology which ensures both privacy protection and high efficiency under complex neural network settings. It employs efficient homomorphic matrix operation to securely execute inference interactively. Furthermore, we have designed and implemented the prototype for PROUD. Finally, we conducted experiments to evaluate the performance using two common datasets. PROUD has been shown to outperform the existing schemes, and can be readily integrated into the practical cloud infrastructure.

\subsection*{Acknowledgments}
This work is partially supported by the NSF under Grant No. CNS-1745894. The authors would like to thank the anonymous reviewers for their constructive comments.



\begin{thebibliography}{10}

\bibitem{liu2017survey}
Weibo Liu, Zidong Wang, Xiaohui Liu, Nianyin Zeng, Yurong Liu, and Fuad~E
  Alsaadi,
\newblock ``A survey of deep neural network architectures and their
  applications,''
\newblock {\em Neurocomputing}, vol. 234, pp. 11--26, 2017.

\bibitem{tran2015learning}
Du~Tran, Lubomir Bourdev, Rob Fergus, Lorenzo Torresani, and Manohar Paluri,
\newblock ``Learning spatiotemporal features with 3d convolutional networks,''
\newblock in {\em Proceedings of the IEEE international conference on computer
  vision}, 2015, pp. 4489--4497.

\bibitem{jiang2018secure}
Xiaoqian Jiang, Miran Kim, Kristin Lauter, and Yongsoo Song,
\newblock ``Secure outsourced matrix computation and application to neural
  networks,''
\newblock in {\em Proceedings of the 2018 ACM SIGSAC Conference on Computer and
  Communications Security}. ACM, 2018, pp. 1209--1222.

\bibitem{tramer2016stealing}
Florian Tram{\`e}r, Fan Zhang, Ari Juels, Michael~K Reiter, and Thomas
  Ristenpart,
\newblock ``Stealing machine learning models via prediction apis,''
\newblock in {\em 25th USENIX Security Symposium (USENIX Security 16)}, 2016,
  pp. 601--618.

\bibitem{juvekar2018gazelle}
Chiraag Juvekar, Vinod Vaikuntanathan, and Anantha Chandrakasan,
\newblock ``Gazelle: A low latency framework for secure neural network
  inference,''
\newblock in {\em 27th USENIX Security Symposium (USENIX Security 18)}, 2018,
  pp. 1651--1669.

\bibitem{gilad2016cryptonets}
Ran Gilad-Bachrach, Nathan Dowlin, Kim Laine, Kristin Lauter, Michael Naehrig,
  and John Wernsing,
\newblock ``Cryptonets: Applying neural networks to encrypted data with high
  throughput and accuracy,''
\newblock in {\em International Conference on Machine Learning}, 2016, pp.
  201--210.

\bibitem{xie2019bayhenn}
Peichen Xie, Bingzhe Wu, and Guangyu Sun,
\newblock ``Bayhenn: Combining bayesian deep learning and homomorphic
  encryption for secure dnn inference,''
\newblock {\em arXiv preprint arXiv:1906.00639}, 2019.

\bibitem{zhang2018gelu}
Qiao Zhang, Cong Wang, Hongyi Wu, Chunsheng Xin, and Tran~V Phuong,
\newblock ``Gelu-net: A globally encrypted, locally unencrypted deep neural
  network for privacy-preserved learning.,''
\newblock .

\bibitem{Yao86}
A.~C. {Yao},
\newblock ``How to generate and exchange secrets,''
\newblock in {\em 27th Annual Symposium on Foundations of Computer Science
  (sfcs 1986)}, Oct 1986, pp. 162--167.

\bibitem{paillier1999public}
Pascal Paillier,
\newblock ``Public-key cryptosystems based on composite degree residuosity
  classes,''
\newblock in {\em International Conference on the Theory and Applications of
  Cryptographic Techniques}. Springer, 1999, pp. 223--238.

\bibitem{cheon2017homomorphic}
Jung~Hee Cheon, Andrey Kim, Miran Kim, and Yongsoo Song,
\newblock ``Homomorphic encryption for arithmetic of approximate numbers,''
\newblock in {\em International Conference on the Theory and Application of
  Cryptology and Information Security}. Springer, 2017, pp. 409--437.

\bibitem{devlin2018bert}
Jacob Devlin, Ming-Wei Chang, Kenton Lee, and Kristina Toutanova,
\newblock ``Bert: Pre-training of deep bidirectional transformers for language
  understanding,''
\newblock {\em arXiv preprint arXiv:1810.04805}, 2018.

\bibitem{carreira2017quo}
Joao Carreira and Andrew Zisserman,
\newblock ``Quo vadis, action recognition? a new model and the kinetics
  dataset,''
\newblock in {\em proceedings of the IEEE Conference on Computer Vision and
  Pattern Recognition}, 2017, pp. 6299--6308.

\bibitem{lecun1998gradient}
Yann LeCun, L{\'e}on Bottou, Yoshua Bengio, Patrick Haffner, et~al.,
\newblock ``Gradient-based learning applied to document recognition,''
\newblock {\em Proceedings of the IEEE}, vol. 86, no. 11, pp. 2278--2324, 1998.

\bibitem{mishra2020delphi}
Pratyush Mishra, Ryan Lehmkuhl, Akshayaram Srinivasan, Wenting Zheng, and
  Raluca~Ada Popa,
\newblock ``Delphi: A cryptographic inference service for neural networks,''
\newblock in {\em 29th USENIX Security Symposium (USENIX Security 20)}, 2020.

\bibitem{goldreich1987play}
O~Goldreich, S~Micali, and A~Wigderson,
\newblock ``How to play any mental game,''
\newblock in {\em Proceedings of the nineteenth annual ACM symposium on Theory
  of computing}, 1987, pp. 218--229.

\bibitem{Gentry:FHE}
Craig Gentry,
\newblock ``Fully homomorphic encryption using ideal lattices,''
\newblock in {\em Proceedings of the Forty-first Annual ACM Symposium on Theory
  of Computing}, New York, NY, USA, 2009, STOC '09, pp. 169--178, ACM.

\bibitem{boyle2017homomorphic}
Elette Boyle, Geoffroy Couteau, Niv Gilboa, Yuval Ishai, and Michele Orr{\`u},
\newblock ``Homomorphic secret sharing: optimizations and applications,''
\newblock in {\em Proceedings of the 2017 ACM SIGSAC Conference on Computer and
  Communications Security}, 2017, pp. 2105--2122.

\bibitem{cryptoeprint:2012:144}
Junfeng Fan and Frederik Vercauteren,
\newblock ``Somewhat practical fully homomorphic encryption,'' Cryptology
  ePrint Archive, Report 2012/144, 2012.

\bibitem{halevi2018faster}
Shai Halevi and Victor Shoup,
\newblock ``Faster homomorphic linear transformations in helib,''
\newblock in {\em Annual International Cryptology Conference}. Springer, 2018,
  pp. 93--120.

\bibitem{halevi2019improved}
Shai Halevi, Yuriy Polyakov, and Victor Shoup,
\newblock ``An improved rns variant of the bfv homomorphic encryption scheme,''
\newblock in {\em Cryptographers’ Track at the RSA Conference}. Springer,
  2019, pp. 83--105.

\bibitem{mohassel2017secureml}
Payman Mohassel and Yupeng Zhang,
\newblock ``Secureml: A system for scalable privacy-preserving machine
  learning,''
\newblock in {\em 2017 IEEE Symposium on Security and Privacy (SP)}. IEEE,
  2017, pp. 19--38.

\end{thebibliography}
\end{document}